\documentclass[review]{elsarticle}

\usepackage{lineno,hyperref}
\usepackage{amsmath}
\usepackage{amssymb}
\usepackage{amsthm}
\usepackage{graphicx}
\usepackage{subfigure}
\usepackage{multirow}
\usepackage{subfigure}
\usepackage{color}
\usepackage{threeparttable}
\usepackage{bm}
\usepackage{stfloats}
\usepackage{bigstrut}
\usepackage{booktabs}
\usepackage{soul}
\UseRawInputEncoding

\journal{Journal of \LaTeX\ Templates}

\begin{document}

\begin{frontmatter}

\title{A Brief Survey on Adaptive Video Streaming Quality Assessment}

\author[a]{Wei Zhou}
\author[b]{Xiongkuo Min}
\author[c]{Hong Li}
\author[c]{Qiuping Jiang\corref{cor1}}
\ead{jiangqiuping@nbu.edu.cn}
\address[a]{Department of Electrical and Computer Engineering, University of Waterloo, Waterloo, ON N2L 3G1, Canada}
\address[b]{Institute of Image Communication and Network Engineering, Shanghai Jiao Tong University, Shanghai 200240, China}
\address[c]{School of Information Science and Engineering, Ningbo University, Ningbo 315211, China}
\cortext[cor1]{Corresponding author}

\begin{abstract}
Quality of experience (QoE) assessment for adaptive video streaming plays a significant role in advanced network management systems. It is especially challenging in case of dynamic adaptive streaming schemes over HTTP (DASH) which has increasingly complex characteristics including additional playback issues. In this paper, we provide a brief overview of adaptive video streaming quality assessment. Upon our review of related works, we analyze and compare different variations of objective QoE assessment models with or without using machine learning techniques for adaptive video streaming. Through the performance analysis, we observe that hybrid models perform better than both quality-of-service (QoS) driven QoE approaches and signal fidelity measurement. Moreover, the machine learning-based model slightly outperforms the model without using machine learning for the same setting. In addition, we find that existing video streaming QoE assessment models still have limited performance, which makes it difficult to be applied in practical communication systems. Therefore, based on the success of deep learned feature representations for traditional video quality prediction, we also apply the off-the-shelf deep convolutional neural network (DCNN) to evaluate the perceptual quality of streaming videos, where the spatio-temporal properties of streaming videos are taken into consideration. Experiments demonstrate its superiority, which sheds light on the future development of specifically designed deep learning frameworks for adaptive video streaming quality assessment. We believe this survey can serve as a guideline for QoE assessment of adaptive video streaming.
\end{abstract}

\begin{keyword}
Quality of experience, video quality assessment, adaptive streaming, performance analysis, deep convolutional neural network, spatio-temporal characteristics.
\end{keyword}

\end{frontmatter}

\section{Introduction}
With the rapid development of network services and mobile devices, streaming related multimedia applications have obtained tremendous growth. The arrival of dynamic adaptive streaming schemes over HTTP (DASH) standard \cite{stockhammer2011dynamic} provides the transition from traditional connection-based video streaming protocols to hypertext transfer protocol (HTTP) adaptive streaming (HAS) protocols which enable flexible deployment, reduced workload, and reliable delivery. In addition, data analysis and artificial intelligence have emerged in a service-driven next-generation wireless communication network \cite{kibria2018big}. Therefore, quality of experience (QoE) for HAS streaming videos has attracted increasing attention in both academia and industry \cite{alreshoodi2013survey}.

As defined by the Telecommunication Standardization Sector of International Telecommunication Union (ITU-T), QoE is the overall acceptability of an application or service as perceived subjectively by the end user \cite{sector2008quality}. Addressing the QoE expectations of end-users is crucial for satisfying the requirements of video streaming services. Since users are the ultimate viewers of streaming videos in most practical applications, subjective QoE assessment \cite{seshadrinathan2010subjective,lin2015mcl,xu2018subjective} is straightforward and reliable for the evaluation of perceptual video streaming quality. Specifically, user studies are conducted, in which a number of subjects are asked to rate the visual quality of different streaming videos. The average of these subjective judgments, i.e. mean opinion score (MOS), is computed for the final quality measurement, which is usually known as the ground truth.

Despite the fact that subjective QoE assessment can deliver the most precise and reliable evaluation, these subjective tests are time-consuming, expensive, and inconvenient. More importantly, they cannot be applied to the real-time multimedia distribution and playback scheduling frameworks. Hence, it is also increasingly desirable to develop highly effective and accurate objective QoE assessment models \cite{chikkerur2011objective,jiang2015supervised,li2016color,zhou2019dual} with low computational complexity for streaming videos, which aims to maintain efficient resource allocation and quality management for existing video services in multimedia delivery systems.

Building an effective QoE assessor for adaptive video streaming faces several foreseeable challenges. First, in addition to compression artifacts, how to evaluate the perceived quality of streaming video is more complicated due to additional network impairments (e.g. initial buffering, playback stalling, etc.) compared to traditional VQA. Second, due to the time-consuming subjective experiments, the established databases for video streaming QoE are relatively small-scale, thus it is difficult to train a deep learning-based model with specific network parameters \cite{kim2017deep}. Third, the quality degradation of video streaming is influenced not only by video spatial characteristics, but also by its temporal attributes.

\begin{figure*}[t]
  \centerline{\includegraphics[width=12cm]{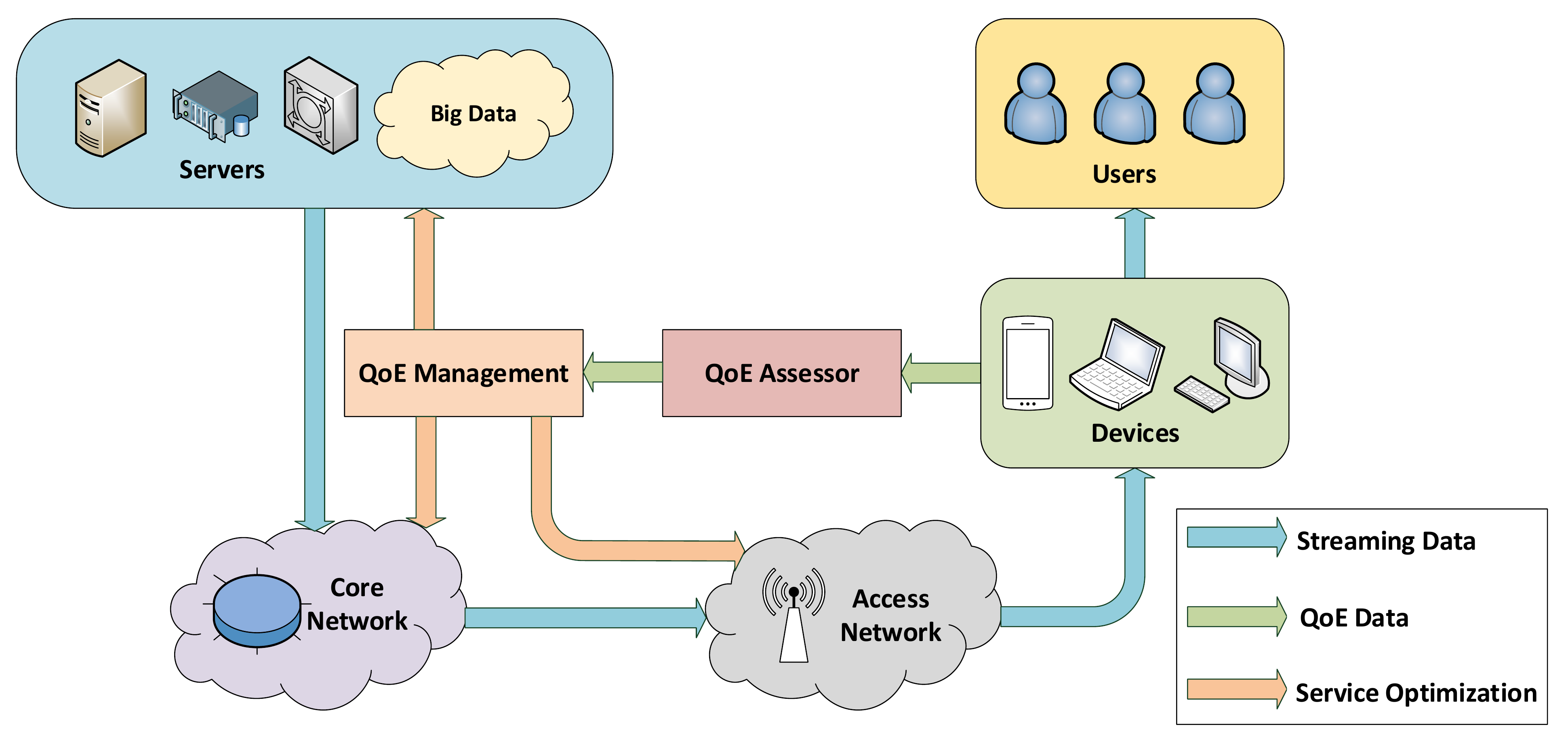}}
  \caption{The important role of QoE assessment in multimedia communication systems. The designed QoE assessor is deployed to take the decoded streaming video data as input, and then predicts perceptual video streaming quality for the service optimization.}
  \centering
\label{fig:fig1}
\end{figure*}

Figure \ref{fig:fig1} shows the important role of QoE assessment in multimedia communication systems. It illustrates that application servers transmit adaptive video streaming data to users through the core network, access network, and terminal devices. Then, the designed QoE assessor is deployed to take relevant information, e.g. the decoded streaming video data as input, and predicts perceptual video streaming quality. Finally, the virtually located QoE management receives the information from the QoE assessor and aims to optimize the network delivery of streaming video data.

Although various surveys have been proposed for the perceptual quality assessment of images/videos \cite{lin2011perceptual,vega2016experimental,zhai2020perceptual}, such reviews for adaptive video streaming quality assessment are relatively scarce. Towards this end, we present an overview of QoE assessment for adaptive video streaming. The main contributions of this paper are summarized in three-fold as follows:

\begin{itemize}
\item We review subjective QoE assessment studies for adaptive video streaming, where the details of various subjective quality databases are described. Different from traditional video subjective QoE assessment, we discuss particular quality factors in these constructed databases.

\item We review existing objective QoE assessment models for adaptive video streaming, including quality-of-service (QoS) driven user QoE assessment, signal fidelity measurement, and hybrid models.

\item We analyze and compare objective QoE assessment models with or without using machine learning for adaptive video streaming. Besides, we exploit the deep feature representations from off-the-shelf DCNN models, which is consistent with spatio-temporal human visual perception. Experiments demonstrate its superior performance, leading to the promising future direction to develop streaming-aware quality prediction frameworks based on DCNN.
\end{itemize}

The structure of this article is as follows. In Section \ref{2} and Section \ref{3}, we review the relevant subjective quality databases and existing objective QoE models for streaming video quality evaluation, respectively. In Section \ref{4}, we analyze and compare different variations of QoE assessment models with or without using machine learning for adaptive video streaming. Additionally, the deep feature representations from off-the-shelf DCNN models show promising results against state-of-the-art methods, which sheds light on the future development of specific deep learning-based quality evaluation frameworks for adaptive video streaming. We conclude the results and discuss some future directions in Section \ref{5}.

\section{Subjective Studies}
\label{2}
The perceptual quality of HAS streaming videos not only suffers from compression distortions, but also degrades due to some streaming-specific issues, such as initial buffering, playback stalling, etc. Up to now, some subjective quality databases have been built for severing as the benchmarks for objective QoE assessment models. Here, we give an introduction to the mainstream publicly available subjective databases for adaptive video streaming during the past decade. The details of publicly available subjective databases for adaptive video streaming can be found in Table \ref{table1}.

\begin{table}[ht]
	\begin{center}
		\caption{Details of Publicly Available Subjective Databases for Adaptive Video Streaming.}
		\label{table1}
		\scalebox{0.65}{
			\begin{tabular}{c|cccc}
				\hline
				Databases & \# of Source Videos & \# of Distorted Videos & \# of Codecs & Viewing Displays \\
                \hline
				LIVEMVQA \cite{moorthy2012video}           & 10 & 300   & 1 & Phone \& Tablet \\
				LIVEQHVS \cite{chen2014modeling}           & 3  & 15    & 1 & HDTV \\
				LIVEMSV \cite{ghadiyaram2014study}         & 24 & 180   & 0 & Phone \\
				WaterlooSQoE-I \cite{duanmu2017quality}    & 20 & 180   & 1 & HDTV \\
				LIVE-NFLX-I \cite{bampis2017study}         & 14 & 112   & 1 & Phone \\
                WaterlooSQoE-II \cite{duanmu2017qoe}       & 12 & 588   & 1 & HDTV \\
                WaterlooSQoE-III \cite{duanmu2018quality}  & 20 & 450   & 1 & HDTV \\
                LIVE-NFLX-II \cite{bampis2021towards}      & 15 & 420   & 1 & HDTV \\
                WaterlooSQoE-IV \cite{duanmu2020assessing} & 5  & 1,350 & 2 & Phone \& HDTV \& UHDTV \\
                \hline
		\end{tabular}}
	\end{center}
\end{table}

The earliest adaptive video streaming subjective database dates back to 2012 when Moorthy et al. proposed the LIVE mobile video quality assessment (LIVEMVQA) database \cite{moorthy2012video}. This database consists of 10 source videos and 300 test videos. The distortion types include H.264 compression, wireless channel packet loss, frame freezes, rate adaptation, and temporal dynamics. In the subjective test, 200 distorted videos evaluated by over 30 subjects on a small phone, as well as 100 distorted videos rated by 17 subjects on a larger tablet device.

The LIVEQHVS \cite{chen2014modeling} contains 3 original reference videos. The length of each video is relatively long, which is 300 seconds duration. The long videos are constructed by concatenating 8 high-quality short video clips. For each source video, 6 quality-varying videos are generated by applying various encoding bitrates of H.264 encoder. Among these 18 distorted videos, 3 of them are used for the training of subjective studies, the remaining 15 quality-varying videos are exploited for testing. These video sequences are displayed to the subjects on a 58-inch high definition television (HDTV) monitor.

The LIVEMSV \cite{ghadiyaram2014study} includes 24 pristine videos with either $1280 \times 720$ pixels or $640 \times 360$ pixels. Since this database focuses on network impairments, other factors such as spatial distortions are minimized. There exist 180 distorted videos produced by all the reference videos with 26 unique hand-crafted stalling events. The subjective quality labels are obtained from 54 subjects, leading to 4,830 human opinions. The viewing display is Apple iPhone 5.

The WaterlooSQoE-I \cite{duanmu2017quality} considers both the compression and playback artifacts. With H.264 encoder, the original source videos are encoded into three bitrate levels which include 500 Kbps, 1,500 Kbps, and 3,000 Kbps. Note that these three bitrate levels are based on commonly available parameters of video transmission over wireless communication networks. In addition, playback issues are also taken into account in this database. To be specific, apart from the introduced compression distortions, a five-second stalling event is then simulated at either the beginning or the middle time point of the encoded video sequences. Therefore, the initial buffering and middle playback stalling streaming videos can be produced by this kind of simulation. In total, this database is made up of 200 videos containing 20 source videos, 60 encoded videos, 60 initial buffering videos, and 60 middle playback stalling videos. A subjective study is conducted to collect ratings of test videos on the HDTV display.

The LIVE-NFLX-I \cite{bampis2017study} is presented to investigate the influence of mixtures from adaptive streaming video artifacts. The database is composed of 14 source video contents and 112 distorted videos obtained by encoding the original videos using H.264 encoder. There are 8 different playout patterns including dynamically changing H.264 compression rates, rebuffering events, and the mixtures of both. The subjective experiment is conducted by 55 subjects on a mobile device. It should be noted that only three reference videos and their corresponding distorted videos are made publicly available in this database.

The WaterlooSQoE-II \cite{duanmu2017qoe} involves 12 source videos, where each video has 8 seconds duration and is further partitioned into 4-second short segments. The short segments are encoded into seven representations with H.264 codec. To simulate quality adaptation events, two consecutive 4-second segments with different representations are concatenated from the same video content, resulting in 588 videos with variations in compression level, spatial resolution, and frame rate. The videos are displayed at their pixel resolution on the HDTV display for subjective quality collections.

The above-mentioned subjective quality databases for adaptive video streaming have a common issue that they are hand-crafted. That is, these databases are far away from real-world streaming video distributions. Therefore, the following recently established databases aim to tackle this problem, including WaterlooSQoE-III \cite{duanmu2018quality}, LIVE-NFLX-II \cite{bampis2021towards}, and WaterlooSQoE-IV \cite{duanmu2020assessing}.

Specifically, the WaterlooSQoE-III \cite{duanmu2018quality} and the LIVE-NFLX-II \cite{bampis2021towards} have 450 and 420 realistic adaptive streaming videos, respectively. Both databases integrate actual network traces to capture realistic network variations. Different realistic adaptive bitrate (ABR) streaming algorithms \cite{li2014probe,yin2015control,akhtar2018oboe} are employed for video delivery. The subjective experiments conducted on the HDTV are used to obtain subjective QoE ratings. The WaterlooSQoE-IV \cite{duanmu2020assessing} provides so far the most comprehensive QoE assessment database which consists of 1,350 subjective-rated streaming videos that are derived from a variety of source video contents, video codecs, network conditions, ABR algorithms, and viewing displays. For example, except for mobile phone and HDTV, the Ultra HDTV (UHDTV) is also applied in subjective studies.

With these subjective quality databases for adaptive video streaming, effective quality labels are provided for designing objective quality assessment algorithms, which facilitates researchers to propose objective quality assessment models that are closer to human visual perception. Besides, we can measure and compare the performance of different adaptive video streaming quality evaluation models on these databases.

\section{Objective Models}
\label{3}
In general, according to the existence of original reference videos, traditional objective Video QoE Assessment (VQA) methods can be classified into full-reference (FR) VQA, reduced-reference (RR) VQA, and no-reference (NR) VQA. Specifically, FR VQA methods \cite{wang2004image,wang2003multiscale,rehman2015display} require the corresponding original reference video. The RR VQA methods assume that a portion of the reference video is available, which can be some parameters extracted from the original content or additional side information added to the test video. The NR VQA algorithms \cite{chen2016hybrid,zhou20163d,chen2018blind} evaluate visual quality without any information from the corresponding original reference video, which are more practical in application scenarios.

In the literature, there have emerged several objective quality assessment models for adaptive video streaming. Apart from the classification method of conventional VQA methods that is based on the information of original reference videos, existing video streaming QoE assessment models can be generally classified into three categories which include QoS driven user QoE assessment \cite{hossfeld2011quantification,rodriguez2012quality}, signal fidelity measurement \cite{wang2004image,wang2003multiscale,rehman2015display}, and hybrid models \cite{duanmu2017quality,bampis2017learning}. Specifically, the QoS driven user QoE assessment exploits the causal relationship between QoS and QoE problems, while the signal fidelity measurement takes the QoE assessment problem from the aspect of signal fidelity. The hybrid models comprehensively consider the QoS driven user QoE assessment and the signal fidelity measurement at the same time. Moreover, solid related work has been done on automatic video streaming QoE assessment models with or without using machine learning techniques. Table \ref{table2} lists the summary of objective QoE assessment models for adaptive video streaming. It should be noted that we only focus on the mentioned QoE assessment models for adaptive video streaming in this paper, other algorithms could be found in \cite{barman2019qoe}.

\begin{table}[ht]
	\begin{center}
		\caption{Summary of objective QoE Assessment Models for Adaptive Video Streaming.}
		\label{table2}
		\scalebox{1.0}{
			\begin{tabular}{c|cc}
				\hline
				Methods & Learning & Types \\
                \hline
				FTW         \cite{hossfeld2011quantification} & No  & QoS driven QoE assessment  \\
				VsQM        \cite{rodriguez2012quality}       & No  & QoS driven QoE assessment  \\
				PSNR                                          & No  & Signal fidelity measurement \\
				SSIM        \cite{wang2004image}              & No  & Signal fidelity measurement \\
				MS-SSIM     \cite{wang2003multiscale}         & No  & Signal fidelity measurement \\
                SSIMplus    \cite{rehman2015display}          & No  & Signal fidelity measurement \\
                SQI         \cite{duanmu2017quality}          & No  & Hybrid model \\
                Video ATLAS \cite{bampis2017learning}         & Yes & Hybrid model \\
                \hline
		\end{tabular}}
	\end{center}
\end{table}

As for QoS driven user QoE assessment, the mapping functions between QoS and QoE problems are usually employed. For example, several QoS models such as FTW \cite{hossfeld2011quantification} and VsQM \cite{rodriguez2012quality} have been proposed by utilizing global rebuffering statistics and the pattern of temporal local content importance. However, the video quality impairment caused by video compression has not been taken into consideration.

For signal fidelity measurement, conventional objective VQA metrics consider human visual perception rather than the simplest peak signal-to-noise ratio (PSNR), such as the structural similarity index (SSIM) \cite{wang2004image}, the multi-scale structure similarity index (MS-SSIM) \cite{wang2003multiscale}, the SSIMplus \cite{rehman2015display}, and so on. However, all of these algorithms are under the assumption that the playback can be exactly controlled. But in the applications of QoE assessment for HAS streaming videos, due to network transmission impairments, these services may suffer from some playback issues which could bring significant quality degradation.

Additionally, hybrid models integrate the scheme of QoS driven user QoE assessment with the scheme of signal fidelity measurement. In \cite{duanmu2017quality}, a unified video streaming QoE assessor without using machine learning named streaming quality index (SQI) has been proposed, which combines FR quality metrics such as SSIM and MS-SSIM with stalling related information. In other words, the impact of compression and stalling are modeled simultaneously. Moreover, in \cite{bampis2017learning}, the machine learning-based model called video assessment of temporal artifacts and stalls (Video ATLAS) has been presented, where a number of QoE-related features, including objective quality features, rebuffering-aware features and memory-driven features, are utilized to predict the perceptual video streaming quality.

\section{Performance Analysis}
\label{4}

\subsection{Performance Evaluation Criteria}
According to the recommendation from the Video Quality Experts Group \cite{video2000final}, we adopt two widely used criteria including the Spearman rank-order correlation coefficient (SROCC) and Pearson linear correlation coefficient (PLCC) to analyze and compare different objective algorithms for video streaming QoE assessment. The SROCC performance is utilized to measure QoE prediction monotonicity, while the PLCC performance is applied to evaluate QoE prediction accuracy.

Moreover, before computing the PLCC performance of objective QoE assessment algorithms, a nonlinear logistic fitting is generally used to map the predicted quality scores to the same scales of subjective quality ratings. Here, higher SROCC and PLCC correlation coefficients indicate better performance and agreement with subjective human quality perception.

\begin{figure*}[t]
  \centerline{\includegraphics[width=12cm]{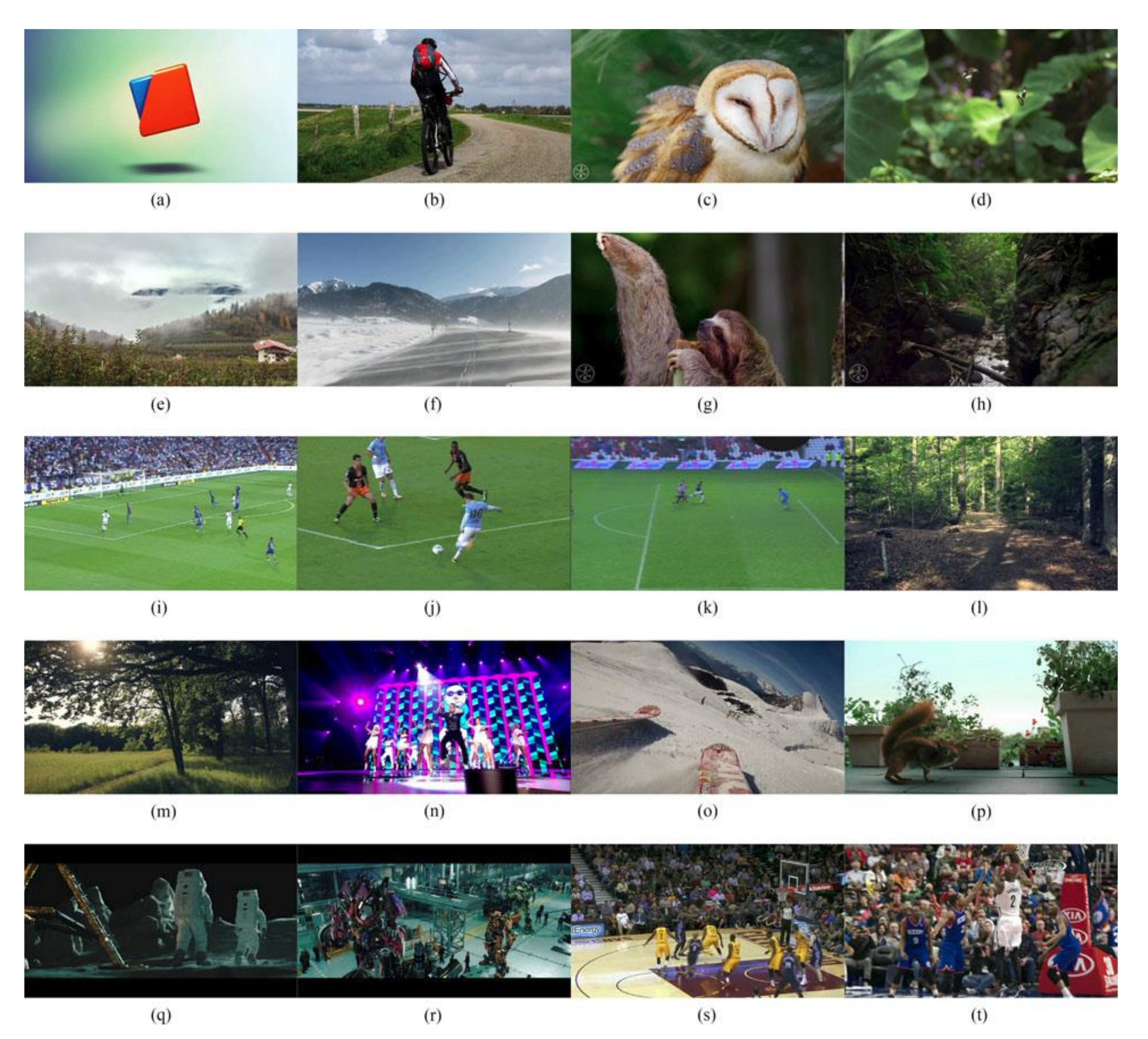}}
  \caption{Video contents in WaterlooSQoE-I database \cite{duanmu2017quality}.}
  \centering
\label{fig:fig2}
\end{figure*}

\subsection{Results of Video Streaming QoE Models}
Considering that existing adaptive video streaming QoE assessment models consist of QoS driven QoE methods, signal fidelity measurement, and hybrid approaches which combine FR quality metrics with rebuffering related features, we conduct experiments on the representative WaterlooSQoE-I database \cite{bampis2017learning}. The reason for choosing the WaterlooSQoE-I database is that it is the first database containing the most abundant video contents and codecs. Figure \ref{fig:fig2} shows that there exist 20 original reference videos consisting of diverse video contents in the WaterlooSQoE-I database \cite{duanmu2017quality}. With these pristine sources, streaming videos impaired from both compression distortions and playback issues can be generated.

\begin{figure*}[t]
  \centerline{\includegraphics[width=12cm]{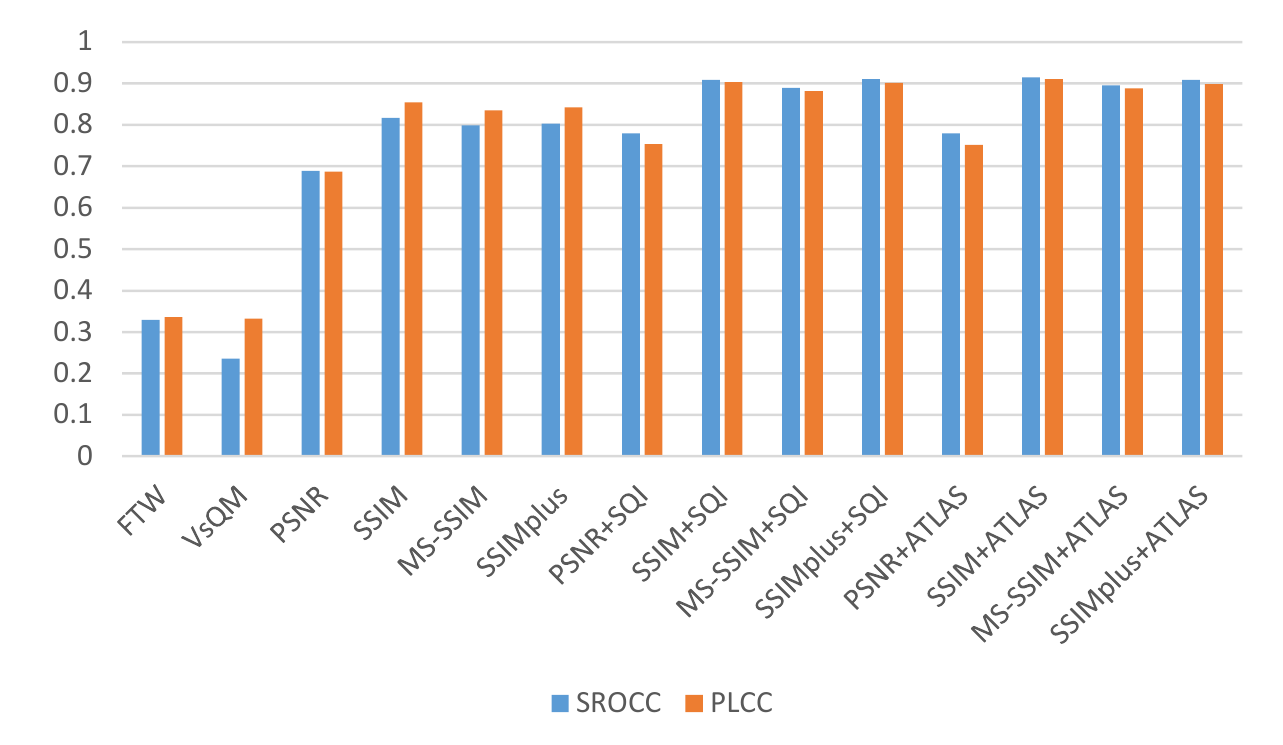}}
  \caption{Performance comparison of existing video streaming QoE assessment models on WaterlooSQoE-I database \cite{duanmu2017quality}.}
  \centering
\label{fig:fig3}
\end{figure*}

We first analyze different variations of objective quality assessment models with or without using machine learning for adaptive video streaming. Figure \ref{fig:fig3} shows the performance comparison of existing video streaming QoE assessment models on WaterlooSQoE-I database \cite{duanmu2017quality}. In this figure, the Video ATLAS is the only machine learning-based method. And the regression model with the best SROCC performance is reported.

In general, as we can see from Figure \ref{fig:fig3}, hybrid models outperform QoS driven QoE models (i.e. FTW and VsQM) as well as the signal fidelity measurement when using the same FR quality metrics. Additionally, the machine learning-based QoE assessment model, namely Video ATLAS, is slightly superior to the SQI model without using machine learning for the same FR quality metrics combination. One possible explanation is that the machine learning-based method learns distortion-related features better directly from streaming videos.

Based on these observations, we can conclude that existing video streaming QoE assessment algorithms still have limited performance, making it difficult to be employed in practical applications. At the same time, deep learning models have been studied to understand the development of human sensory cortical processing \cite{yamins2016using}. Besides, deep convolutional neural network (DCNN) has been applied to perceptual quality assessment, which demonstrates the remarkable ability of DCNN to learn discriminative features for addressing this challenging task \cite{lopez2018deep,zhou2018stereoscopic,zhou2020deep}. However, to the best of our knowledge, there is no similar research work about deep learning solutions to hybrid NR video streaming QoE assessment based on spatio-temporal visual content features in streaming videos. Thus, we apply a simplified framework to tackle the adaptive video streaming quality assessment, where we only exploit distorted streaming videos without referring to the corresponding source videos. That is, this method belongs to the category of no-reference video streaming QoE models, which is more practical in real applications.


Specifically, considering that a streaming video sequence is composed of many video frames, we first extract multiple distorted video frames from streaming video sequences. Apart from the spatial characteristics of different distorted video frames, note that a video sequence is a set of consecutive video frames which contain a variety of motion attributes. In other words, the temporal variation of video contents could affect the visual perception of the HVS to a certain degree. Thus, compared to image quality prediction, video streaming QoE assessment is more complex due to the additional temporal quality variation. We then utilize frame difference maps to take the temporal factor into account, which are simply defined as the difference between adjacent video frames.

The two pre-trained DCNN models take distorted video frames and frame difference maps as inputs to extract the 2,048-dim features from the pool5 layer. Note that the two employed ResNet50 architectures have the same configuration and share weights with each other. Then, the concatenation of the extracted 2,048-dim features constitute a 4,096-dim feature vector. Finally, the well-known regression model (i.e. SVR) is applied to map the 4,096-dim feature vector into the ultimate perceptual quality score for each streaming video. In addition, it should be noted that the used database is randomly divided into 80\% for training and the remaining 20\% for testing. We perform 1,000 iterations of cross correlation, and then give the median SROCC and PLCC values as the final measurement.

\begin{figure*}
  \centerline{\includegraphics[width=12cm]{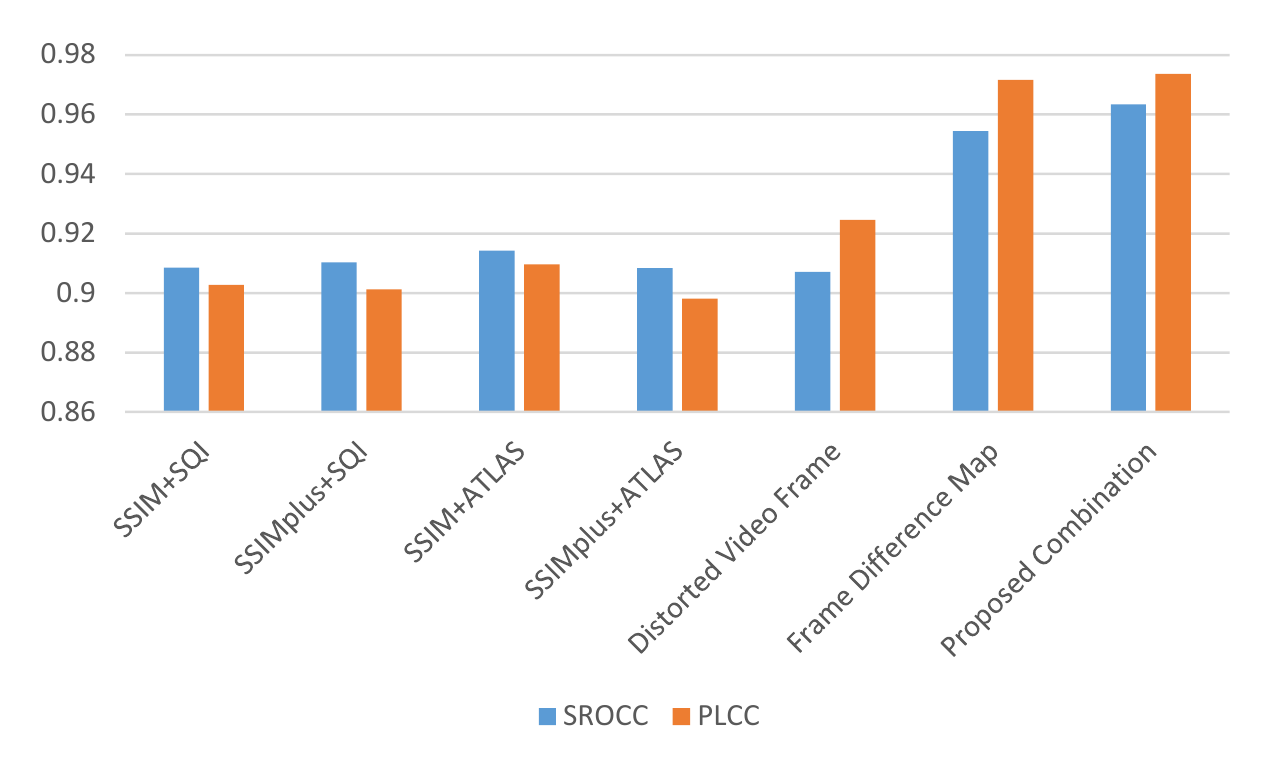}}
  \caption{Performance comparison of the extended video streaming QoE assessment model and state-of-the-arts on WaterlooSQoE-I database \cite{duanmu2017quality}.}
  \centering
\label{fig:fig5}
\end{figure*}

Figure \ref{fig:fig5} shows the performance comparison with state-of-the-art video streaming QoE models. Note that we choose the top two algorithms shown in Figure \ref{fig:fig3} to be compared, which include no machine learning (i.e. SSIM+SQI and SSIMplus+SQI) and machine learning-based (i.e. SSIM+ATLAS and SSIMplus+ATLAS) video streaming QoE assessment models. We denote the deep learning models as ``Distorted Video Frame'', ``Frame Difference Map'', and ``Proposed Combination'', in which we separately apply the deep learned features from distorted video frames, frame difference maps, and the combination of distorted video frames and frame difference maps. As shown in this figure, the deep learning models perform better than the other methods. Moreover, only using frame difference maps outperforms that of only using distorted video frames. One possible explanation may be that the temporal motion attributes have more impact on the perceptual quality of streaming videos compared with that of the spatial texture characteristics. Additionally, the combination of spatial and temporal features outperforms either the spatial feature or the temporal feature alone, which further verifies the significance of spatio-temporal human visual perception in adaptive video streaming.

\begin{figure}
  \centerline{\includegraphics[width=6cm]{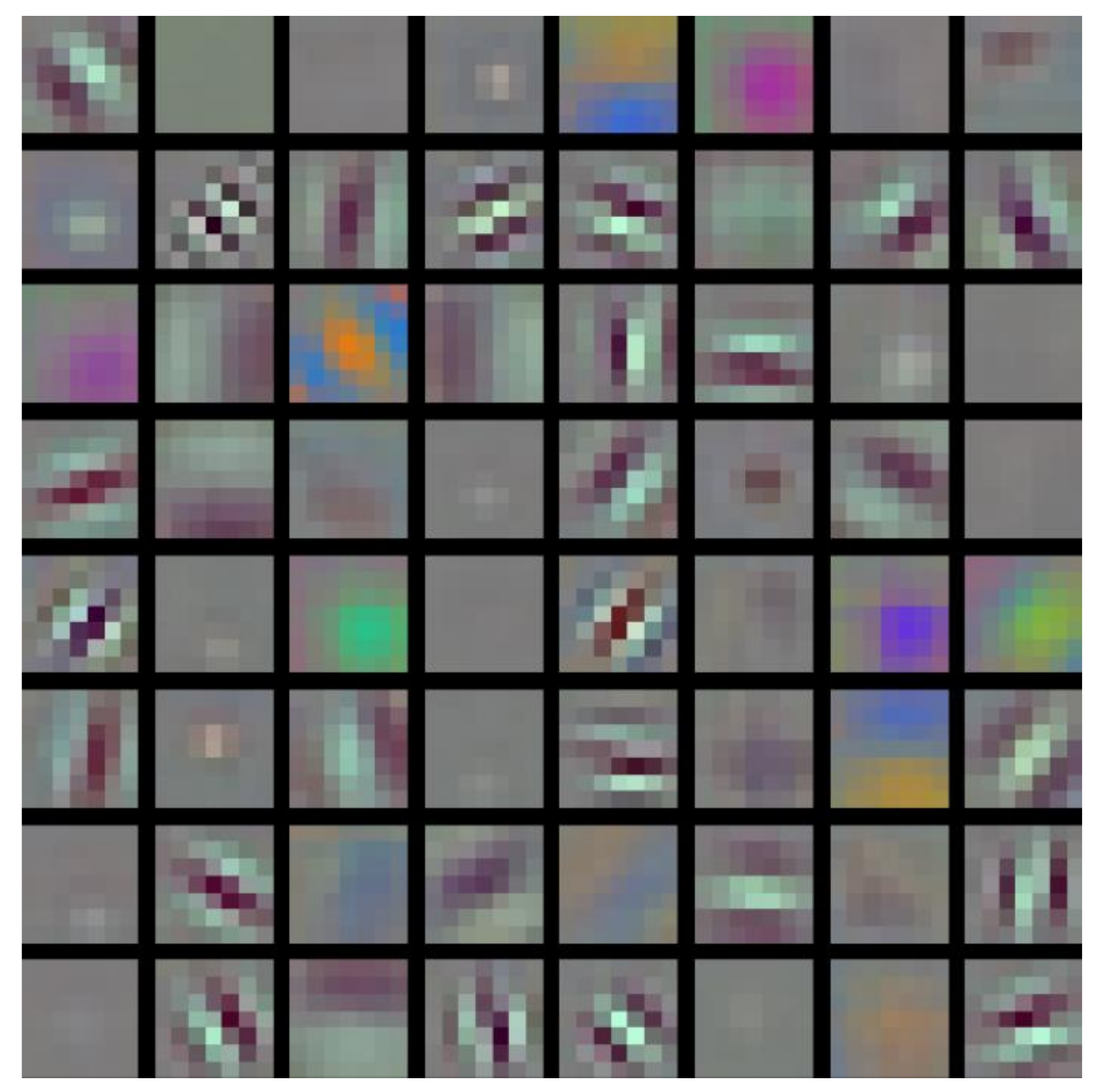}}
  \caption{Visualization of learned kernels in the first convolutional layer for the pre-trained ResNet50 network. The learned kernels capture various intrinsic image texture patterns.}
  \centering
\label{fig:fig6}
\end{figure}

To reveal the discriminative information from deep learned features, Figure \ref{fig:fig6} presents the visualization of learned kernels in the first convolutional layer for the pre-trained ResNet50 network. It should be noted that the ResNet50 model is pre-trained on a large-scale dataset with diverse image contents, namely ImageNet \cite{deng2009imagenet}. Therefore, we can see that the learned kernels can capture intrinsic image texture patterns. In other words, the pre-trained ResNet50 model has a promising ability to represent discriminative features for quality assessment, which sheds light on the future development of specifically designed deep learning models for adaptive video streaming quality evaluation.

\section{Conclusion and Future Directions}
\label{5}
In this paper, we present a brief survey of QoE assessment for adaptive video streaming. First, the QoE assessor plays a vital role in multimedia communication systems. Considering complex characteristics of streaming videos, many challenges are involved in the perceptual quality prediction task. We then review both subjective studies and objective models for adaptive video streaming quality assessment. Finally, we conduct comparisons of existing state-of-the-art QoE assessment models for streaming videos, with or without using machine learning techniques. The performance analysis shows that hybrid models outperform QoS driven QoE models and the signal fidelity measurement when using the same FR quality metrics. Furthermore, the machine learning-based QoE assessment model is demonstrated slightly superior to the model without using machine learning. However, these approaches still have limited performance for video streaming QoE assessment.

Additionally, we apply the deep feature representations from off-the-shelf DCNN models based on spatio-temporal human visual perception, which can deliver promising results. This demonstrates that specific deep learning frameworks for QoE assessment of adaptive video streaming should be addressed in the future. Furthermore, more comprehensive investigation about adaptive video streaming quality assessment could be considered, where the design of immersive 3D/stereoscopic video streaming QoE assessment methods based on deep neural networks is another research direction. For 3D video streaming QoE assessment, except for video quality, more quality dimensions should be taken into consideration, e.g. depth perception and visual comfort.

\bibliography{references}

\end{document}